# A Low-Frequency and Refinement Stable Impedance Boundary Condition EFIE

Alexandre Dély, Francesco P. Andriulli, and Kristof Cools.

**Abstract**

In this contribution, a discretization of the IBC-EFIE is introduced that (i) yields the correct solution at arbitrarily small frequencies, (ii) requires for its solution a number of matrix vector products bounded as the frequency tends to zero and as the mesh density increases. The low frequency stabilization is based on a projector-based discrete Helmholtz splitting, rescaling, and recombination that depends on the low frequency behavior of both the EFIE operator and the surface impedance condition. The dense mesh stabilization is a modification of the Perfect Electric Conductor operator preconditioning approach taking into account the effect on the singular value spectrum of the IBC term.

## I. Introduction

The scattering of time-harmonic electromagnetic waves by non-penetrable objects can be efficiently modeled by the Electric Field Integral Equation (PEC-EFIE) [1]. In this equation, the unknown is the tangential trace of the magnetic field on the boundary of the scatterer. For perfectly conducting objects this trace equals the current induced at the surface of the scatterer and the EFIE is equivalent to the statement that the tangential components of the scattered field generated by this induced current negate the incident field at the surface of the scatterer.

The EFIE can be generalized to model scattering by a much wider class of non-penetrable objects. This is classically obtained by leveraging impedance boundary conditions (IBCs) [2]. These conditions, which will be starting point for this paper, enforce a relationship between between tangential components of electric and magnetic fields, i.e. $\boldsymbol{e}_t = z\hat{\boldsymbol{n}} \times \boldsymbol{h}_t$, or equivalently $\boldsymbol{m} = -z\hat{\boldsymbol{n}} \times \boldsymbol{j}$, where $(\boldsymbol{m}, \boldsymbol{j})$ are the equivalent electric and magnetic surface currents, and $z$ is a so-called surface impedance. When the IBC condition is combined with the representation formulas, it is possible to obtain a first kind equation which is usually known in literature as the Impedance Boundary Condition EFIE (IBC-EFIE) [3]. This equation includes an extra term that takes into account the surface impedance condition. This technique is widespread and is known to provide accurate models of scattering in a wide variety of scenarios. This notwithstanding several recent contributions have sensibly advanced the original integral equation

This work was supported in part by the French DGA agency in the framework of grants for doctoral theses in cooperation between France and United Kingdom.

approach by proposing combined field formulations [4], discretizations based on dual elements [5], self-dual schemes [6], generalized impedance boundary conditions [7], [8].

It is well known that the standard EFIE operator suffers from ill-conditioning when the frequency is low or the discretization density is high. This is often referred to as low-frequency and dense-mesh breakdown, respectively (see [9] and references therein). Although the remarkable advancement in the topic, all IBC formulations currently available are plagued by at least one of the two breakdowns. The concurrent solution of both breakdowns for the IBC-EFIE will be the aim of this paper.

Recently, by leveraging quasi-Helmholtz projector decomposition techniques, the standard EFIE has been rendered numerically stable and accurate at very low frequencies approaching and including zero [10]. On one hand this allow the EFIE to be used in multi-scale problems, without the necessity to couple the EFIE with a dedicated eddy current modeling tool on the boundary of the low and extremely low frequency regions. On the other hand, the EFIE system can be preconditioned using Calderon preconditioning techniques, resulting in linear systems whose condition numbers are virtually independent of the mesh size. This again is a prerequisite for the modeling of scattering by multi-scale problems and in addition allows for the recovery of the exact solution up to arbitrary accuracy.

These methods unfortunately do not trivially carry over to the IBC-EFIE. The reason is that the presence of the term stemming from the IBC affects the scaling of the system matrix blocks in a Helmholtz decomposed basis and the asymptotic behavior of the singular values as the mesh parameter tends to zero. In this contribution, a new formulation of the IBC-EFIE is introduced that (i) yields the correct solution at arbitrarily small frequencies, (ii) has a condition number bounded as the frequency tends to zero, and (iii) has a condition number bounded as the mesh parameter goes to zero. The low frequency stabilization is based on a discrete Helmholtz splitting, rescaling, and recombination that depends on the low frequency behavior of both the EFIE operator and the surface impedance condition. The dense grid stabilization is a modification of the PEC Calderon approach taking into account the effect on the singular value spectrum of the IBC term.

This paper is organized as follow: in Section II the notation is set and the IBC-EFIE is constructed based on the representation theorem. The discretization of the IBC-EFIE leveraging both primal and dual finite element spaces is revisited. In Section III the low frequency behavior of the Helmholtz components of the solution of the IBC-EFIE is studied by considering its solution on a spherical surface in terms of vector spherical harmonics. From this analysis rescaling operators are constructed resulting in a system that is well conditioned up to arbitrarily low frequencies and that is not susceptible to numerical cancellation in the presence of limited machine precision or quadrature errors. In Section IV, the condition number at fixed frequency as the mesh parameter tends to zero is studied. As in Section III, the analysis is first performed for the solution of the IBC on spherical surfaces and then generalized by considering the behavior of the system matrix blocks in a discrete Helmholtz decomposed basis. Based on the conclusions of this analysis, a Calderon-like approach is proposed an the details of its construction elucidated. Finally in Section VI numerical results are presented demonstrating the claims made in this introduction. Both benchmark examples and real life scenarios are considered.

## II. Notation and Background

Consider a domain $\Omega$ with boundary $\Gamma$. The domain is embedded in a medium characterized by a permittivity $\epsilon$ and permeability $\mu$, corresponding to an impedance $\eta = \frac{\mu\omega}{k}$ and wave number $k = \omega\sqrt{\epsilon\mu}$. On $\Gamma$ the exterior normal is denoted $\hat{\boldsymbol{n}}$. Solutions $(\boldsymbol{e}, \boldsymbol{h})$ to Maxwell's equations in the exterior domain $\mathbb{R}^3 \setminus \Omega$ that in addition fulfill Sommerfeld's radiation condition are also solutions to the following set of equations:

$$\begin{pmatrix} \frac{1}{2} + K & \eta T \\ -\frac{1}{\eta} T & \frac{1}{2} + K \end{pmatrix} \begin{pmatrix} \boldsymbol{e} \times \hat{\boldsymbol{n}} \\ \hat{\boldsymbol{n}} \times \boldsymbol{h} \end{pmatrix} = \begin{pmatrix} \boldsymbol{e}^i \times \hat{\boldsymbol{n}} \\ \hat{\boldsymbol{n}} \times \boldsymbol{h}^i \end{pmatrix} \tag{1}$$

where

$$T(\boldsymbol{j}) = ikT_s + \frac{1}{ik}T_h \tag{2}$$

$$= -ik\hat{\boldsymbol{n}} \times \int_\Gamma \frac{e^{-ikR}}{4\pi R}\boldsymbol{j}(\boldsymbol{r}')d\boldsymbol{r}'$$

$$\frac{1}{ik}\hat{\boldsymbol{n}} \times \nabla \int_\Gamma \frac{e^{-ikR}}{4\pi R}\nabla' \cdot \boldsymbol{j}(\boldsymbol{r}')d\boldsymbol{r}' \tag{3}$$

and

$$K(\boldsymbol{j}) = \hat{\boldsymbol{n}} \times \text{p.v.} \int_\Gamma \nabla \frac{e^{-ikR}}{4\pi R} \times \boldsymbol{j}(\boldsymbol{r}')d\boldsymbol{r}', \tag{4}$$

are the single and double layer boundary integral operators for the Maxwell system. Here, $R = |\boldsymbol{r} - \boldsymbol{r}'|$ and p.v. denotes the Cauchy principal value of the integral.

This set of equations (1) does not have a unique solution. In fact, given a solution, an infinite set of other solutions can be constructed by adding the fields generated in free space by any source configuration contained in $\Omega$. In order to single out a unique solution, additional constraints need to be enforced.

The following *impedance boundary condition* is assumed to hold

$$\boldsymbol{m} = -z\hat{\boldsymbol{n}} \times \boldsymbol{j}, \tag{5}$$

where $z$ is a complex number. Using (5) in the first equation of (1) results in the Impedance Boundary Condition EFIE (IBC-EFIE)

$$S(\boldsymbol{j}) = \eta T\boldsymbol{j} - \left(\frac{1}{2} + K\right) z\left(\hat{\boldsymbol{n}} \times \boldsymbol{j}\right) = \boldsymbol{e}^i \times \hat{\boldsymbol{n}} \tag{6}$$

### A. Discretization strategy

The discretization strategy proposed in [4] is briefly summarized here to fix the notation. In order to discretize (6), the surface $\Gamma$ is approximated by a flat faceted triangular mesh $\mathcal{T}$ comprising $V$ vertices and $F$ faces. On this mesh the basis of Rao-Wilton-Glisson (RWG) functions [11], $\boldsymbol{f}_m$, $m = 1, ..., N$, is constructed to discretize the current $\boldsymbol{j}$. The RWG basis functions are normalized such that the integral of their normal component over the defining edge equals one (this differs from a factor edge length from the definition found in [11]). The Buffa-Christiansen (BC) functions introduced in [12], $\boldsymbol{g}_m$, $m = 1, ..., N$ are used to discretize the current $\boldsymbol{m}$. These functions are linear combinations of RWG basis functions defined on to the barycentric refinement of $\mathcal{T}$.

Substituting the approximations $\boldsymbol{j} \approx \sum_{n=1}^{N} \mathsf{j}_n \boldsymbol{f}_n$ and $\boldsymbol{m} \approx \sum_{n=1}^{N} \mathsf{m}_n \boldsymbol{g}_n$ and testing (5) with $(\boldsymbol{f}_m)_{m=1}^{N}$ gives [4]

$$\mathsf{m} = -z\mathbf{G}_{mix}^{-1}\mathbf{G}\mathsf{j} \tag{7}$$

where $(\mathbf{G}_{mix})_{i,j} = \langle \hat{\boldsymbol{n}} \times \boldsymbol{f}_i, \boldsymbol{g}_j \rangle$, $(\mathbf{G})_{i,j} = \langle \boldsymbol{f}_i, \boldsymbol{f}_j \rangle$ and $\langle ., . \rangle$ denotes the $L^2(\Gamma)$ inner product. Then the equation (6) is discretized as

$$\mathbf{S}\mathsf{j} = \left( \eta \mathbf{T} - z \left( \mathbf{K} + \frac{1}{2}\mathbf{G}_{mix} \right) \mathbf{G}_{mix}^{-1}\mathbf{G} \right) \mathsf{j} = \mathsf{V} \tag{8}$$

where $(\mathbf{T})_{i,j} = \langle \hat{\boldsymbol{n}} \times \boldsymbol{f}_i, T\boldsymbol{f}_j \rangle$, $(\mathbf{K})_{i,j} = \langle \hat{\boldsymbol{n}} \times \boldsymbol{f}_i, K\boldsymbol{g}_j \rangle$ and $(\mathsf{V})_m = \langle \hat{\boldsymbol{n}} \times \boldsymbol{f}_m, \boldsymbol{e}^i \times \hat{\boldsymbol{n}} \rangle$.

The coefficients of both RWG and BC functions allow for a discrete Helmholtz decomposition [13]. First, define the connectivity matrices $\mathbf{\Lambda} \in \mathbb{R}^{N\times V}$ and $\mathbf{\Sigma} \in \mathbb{R}^{N\times F}$

$$\mathbf{\Sigma}_{mi} = \pm 1 \quad \text{if edge } m \text{ leaves/arrives at vertex } i, \; 0 \text{ otherwise} \tag{9}$$

$$\mathbf{\Lambda}_{mj} = \pm 1 \quad \text{if edge } m \text{ is on the boundary of face } j \text{ clockwise, countercw, } 0 \text{ otherwise} \tag{10}$$

The space of RWG coefficients $\{\mathsf{j} \in \mathbb{C}^N\}$ is now split into the direct sum of two subspaces. The subspace $\operatorname{Im}\mathbf{\Sigma}$ of RWG stars, which for convenience will also be denoted $\Sigma$, and its $l^2(N)$ orthogonal complement $\Lambda H$ (note that for simply connected surfaces $\Lambda H = \Lambda = \operatorname{Im}\mathbf{\Lambda}$. Here, $l^2(N)$ is $\mathbb{C}^N$ endowed with the Euclidean inner product.

The condition number of equation (8) grows when the frequency decreases or the discretization density increases. These effects are inherited from the low-frequency breakdown ($k \to 0$) and the dense mesh breakdown ($h \to 0$) of the standard EFIE operator. This paper focuses on the solution of these breakdowns for the IBC-EFIE.

## III. Analysis and Regularization of the Low-Frequency Breakdown of IBC-EFIE

To gain initial insight in the low-frequency behavior of the IBC-EFIE, a Mie series analysis based on an expansion in vector spherical harmonics is presented in this section. Although this strategy does not strictly apply to the general case, its findings can be generalized to arbitrary smooth geometries. Define the vector spherical harmonics

$$\boldsymbol{X}_{lm}(\theta,\varphi) = \frac{a}{i\sqrt{l(l+1)}} \hat{\boldsymbol{n}} \times \nabla Y_{lm}(\theta,\varphi) \tag{11}$$

$$\boldsymbol{U}_{lm}(\theta,\varphi) = \hat{\boldsymbol{n}} \times \boldsymbol{X}_{lm}(\theta,\varphi) \tag{12}$$

with $Y_{lm}$ the scalar spherical harmonics and $a$ the sphere radius. The vector spherical harmonics are singular vectors of both $T$ and $K$ operators. In particular, it holds [14] that

$$S(\boldsymbol{U}) = [(\;\;\eta H_l'(ka) + ziH_l(ka))J_l'(ka)]\,\boldsymbol{X} \tag{13}$$

$$S(\boldsymbol{X}) = [(-\eta H_l(ka) + ziH_l'(ka))J_l(ka)]\,\boldsymbol{U} \tag{14}$$

where $H_l$, $J_l$, $H_l'$ and $J_l'$ denote the Riccati-Hankel, Riccati-Bessel and their derivatives, respectively. The asymptotic behavior of these special functions for $k \to 0$ (see, for example, [15]), assuming that $z = o(k^{-1})$, for $k \to 0$, results in the following asymptotic estimates

$$S(\boldsymbol{U}) = \left[\underset{k\to 0}{O}\left(\frac{\eta l}{ikb}\right)\right]\boldsymbol{X} = \left[\underset{k\to 0}{O}\left(\frac{\eta l\sigma}{ik}\right)\right]\boldsymbol{X} \tag{15}$$

$$S(\boldsymbol{X}) = \left[\underset{k\to 0}{O}\left(\frac{a\eta ik}{l} + z\right)\right]\boldsymbol{U} = \left[\underset{k\to 0}{O}\left(\frac{ik\eta\tau}{l} + z\right)\right]\boldsymbol{U}. \tag{16}$$

Here $\sigma$ is the smallest singular value of $T_h$ and $\tau$ is the largest singular value of $T_s$. Stating these asymptotic expressions allows to generalize the conclusions in this section to arbitrary geometries. In order to facilitate tracking the dimensionality of various quantities the length scales $b = 1/\sigma$ and $a = \tau$ are introduced (for the symmetric case of the sphere they happen to be the same).

From this it follows that the condition number of (6), for a fixed number of terms $l_{\max}$ in the Mie series expansion scales as

$$\operatorname{cond}(S) = \underset{k\to 0}{O}\left(\frac{\eta l_{\max}}{ikb\left(z + \frac{ia\eta k}{l_{\max}}\right)}\right). \tag{17}$$

This results in a low-frequency breakdown every time $z = o(k^{-1})$, which holds in all cases of practical interest. This section is concerned with the low frequency behavior of the condition number so in the following computations global constant factors and the truncation point $l_{\max}$ of the Mie series will be omitted. It should also be noted that when $z = 0$ we recover the quadratic-in-frequency growth of the condition number which characterizes the classical EFIE for metallic surfaces.

In addition to conditioning problems, low-frequency regimes are often associated to numerical cancellations in the solution current. This is the case for the standard EFIE [10]. In the following we will assess the problem for IBC-EFIE in the case of plane wave incidence. For a plane wave the following scalings for the right-hand-side hold:

$$\langle \boldsymbol{X}, \hat{n} \times \boldsymbol{E}^i \rangle = O(1) \tag{18}$$

$$\langle \boldsymbol{U}, \hat{n} \times \boldsymbol{E}^i \rangle = O(ikd), \tag{19}$$

where $d$ is a quantity expressed in units of length that depends on the geometry of the scatterer and the incident field. These scalings, combined with (5), (15), and (16), result in the following scalings for the current

$$\langle \boldsymbol{X}, \boldsymbol{j} \rangle = \underset{k\to 0}{O}\left(\frac{ikd}{z + i\eta ka}\right) \tag{20}$$

$$\langle \boldsymbol{U}, \boldsymbol{j} \rangle = \underset{k\to 0}{O}\left(\frac{ikb}{\eta}\right) \tag{21}$$

$$\langle \boldsymbol{X}, \boldsymbol{m} \rangle = \underset{k\to 0}{O}\left(\frac{zikb}{\eta}\right) \tag{22}$$

$$\langle \boldsymbol{U}, \boldsymbol{m} \rangle = \underset{k\to 0}{O}\left(\frac{zikd}{z + i\eta ka}\right). \tag{23}$$

The analysis above shows that, differently from the standard EFIE, for the IBC-EFIE low frequency breakdown does not always occur. The presence of low frequency cancellation depends on the variation of the impedance $z$ with frequency. In fact the above scalings shows that every time the impedance will go faster to zero than $k$ at low frequencies, cancellation will occur in the current.Whenever the impedance goes to a finite non-zero value $0 < z(0) < \infty$ for $k \to 0$ both solenoidal and irrotational components of the solution will have the same frequency scaling (in fact, both will tend to zero proportional to $k$): no numerical cancellation will occur at low frequency.

### A. Solution of the low-frequency problems of the IBC-EFIE

This subsection will show how to solve both low-frequency breakdown and low-frequency current cancellation issues occurring in the solution of the IBC-EFIE. Our strategy will be based on the quasi-Helmholtz projectors introduced in [10] for the standard EFIE. We briefly define these projectors here for the sake of completeness. Given an RWG coefficient vector $\mathsf{j} \in \mathbb{C}^N$ it holds that

$$\mathsf{j} = \mathbf{P}^{\Sigma}\mathsf{j} + \mathbf{P}^{\Lambda H}\mathsf{j}, \tag{24}$$

with $\mathbf{P}^{\Sigma} = \mathbf{\Sigma}\left(\mathbf{\Sigma}^T\mathbf{\Sigma}\right)^{+}\mathbf{\Sigma}^T$ and $\mathbf{P}^{\Lambda H} = \mathbf{I} - \mathbf{P}^{\Sigma}$. Note that the action of the pseudo inverse on any array can be computed in linear complexity using off-the-shelf algorithms [13]. The splitting is orthogonal in the space of coefficients $l^2(N)$. We define the rescaling operators

$$\mathbf{M}_1 = \mathbf{P}^{\Sigma} + \frac{1}{ikd}\mathbf{P}^{\Lambda H} \tag{25}$$

$$\mathbf{M}_2 = ikb\mathbf{P}^{\Sigma} + \frac{\eta ikd}{z + i\eta ka}\mathbf{P}^{\Lambda H} \tag{26}$$

The following low-frequency regularization for the IBC-EFIE is proposed:

$$\mathbf{M}_1\mathbf{S}\mathbf{M}_2\mathsf{Y} = \mathbf{M}_1\mathsf{V} \tag{27}$$

where, as before, $\mathbf{S} = \eta\mathbf{T} - z(\mathbf{K} + \frac{1}{2}\mathbf{G}_{mix})\mathbf{G}_{mix}^{-1}\mathbf{G}$ and where, $\mathsf{j}$ is then retrieved as $\mathsf{j} = \mathbf{M}_2\mathsf{Y}$.

The analysis based on spherical harmonics described in the previous section suggests that (27) is indeed immune from low-frequency problems. As mentioned in the derivation, the length scales $a, b, d$ depend on the geometry and the incident field. As the asymptotic behavior does not depend upon their choice they are put to 1 meter in the numerical experiments in Section VI. Carefully choosing them can further optimize the asymptotic condition number potentially resulting in a reduced number of iterations for the solution of the system.

Applying $\mathbf{M}_1$ and $\mathbf{M}_2$ on the equation (8) results in a system that is immune to the low frequency breakdown. This can be seen in the following way. Define

$$\mathbf{T} = ik\mathbf{T}_s + \frac{1}{ik}\mathbf{T}_h, \tag{28}$$

$$\mathbf{K}^+ = (\mathbf{K} + \frac{1}{2}\mathbf{G}_{mix})\mathbf{G}_{mix}^{-1}\mathbf{G}, \tag{29}$$

then the low frequency behavior of the frequency regularized IBC-EFIE is

$$\begin{aligned}\mathbf{M}_1\mathbf{S}\mathbf{M}_2 &= \mathbf{P}^{\Sigma}(-bk^2\mathbf{T}_s + b\eta\mathbf{T}_h + bikz\mathbf{K}^{+})\mathbf{P}^{\Sigma}\\ &+ \mathbf{P}^{\Sigma}(-\frac{b\eta^2k^2}{z+ik\eta a}\mathbf{T}_s + \frac{d\eta ikz}{z+ik\eta a}\mathbf{K}^{+})\mathbf{P}^{\Lambda H}\\ &+ \mathbf{P}^{\Lambda H}(\frac{ik\eta b}{d}\mathbf{T}_s + \frac{bz}{d}\mathbf{K}^{+})\mathbf{P}^{\Sigma}\\ &+ \mathbf{P}^{\Lambda H}(\frac{ik\eta}{z+ik\eta a}\mathbf{T}_s + \frac{\eta z}{z+ik\eta a}\mathbf{K}^{+})\mathbf{P}^{\Lambda H}\\ &= b\eta\mathbf{P}^{\Sigma}\mathbf{T}_h\mathbf{P}^{\Sigma} + \frac{\eta z}{z+ik\eta a}\mathbf{P}^{\Lambda H}\mathbf{K}^{+}\mathbf{P}^{\Sigma}\\ &+ \frac{\eta z}{z+ik\eta a}\mathbf{P}^{\Lambda H}\mathbf{K}^{+}\mathbf{P}^{\Lambda H} + \mathop{O}_{k\to 0}(k)\end{aligned} \tag{30}$$

It can be read off from this asymptotic estimate that an off-diagonal block tends to zero while the diagonal blocks tend to constant matrices independent of $k$. The condition number, and in turn the number of iterations required to solve the discrete IBC-EFIE neither depends on $k$. Otherwise said, (27) is immune from low-frequency breakdown.

## IV. Analysis and Regularization of the Dense Mesh Breakdown of IBC-EFIE

As in the case of the low-frequency breakdown, spherical harmonics can be used to analyze the conditioning problem of the IBC-EFIE when the discretization density increases (i.e., $h \to 0$). In fact, equation (13) and (14), together with the asymptotic scalings of spherical harmonics and their derivatives for high order provides

$$S(\boldsymbol{U}) = \left[\mathop{O}_{l\to\infty}\left(\frac{i\eta l}{kb}\right)\right]\boldsymbol{X} \tag{31}$$

$$S(\boldsymbol{X}) = \left[\mathop{O}_{l\to\infty}\left(\frac{ia\eta k}{l} + z\right)\right]\boldsymbol{U}. \tag{32}$$

In a vector harmonics expansion truncated at $l$ the number of degrees of freedom is $O(l^2)$. Comparing this to the number of degrees of freedom in a boundary element method discretization $O\left(a^2/h^2\right)$ leads to the approximate relationship $l = a/h$ and thus the estimate for the condition number of the BEM system in terms of the mesh parameter $h$

$$\mathrm{cond}\,(S) = \mathop{O}_{h\to 0}\left(\frac{i\eta}{kbh(z+i\eta kah)}\right) = \mathop{O}_{h\to 0}\left(\frac{1}{h}\right). \tag{33}$$

where the last passage is obtained under the hypothesis that $z \neq 0$. It should be noted that differently from the standard EFIE, that shows a conditioning which is $O(1/h^2)$, the IBC-EFIE condition number is only linearly growing with the inverse of the mesh parameter $h$. This is so because the branch of the spectrum associated to the (compact) operator $T_s$ is dominated in the IBC-EFIE by the presence of the identity (which is absent in the standard EFIE).

### *A. Solution of the dense mesh problems of the IBC-EFIE*

A further addition to the new formulation in (27) will result in an equation that is immune from both low frequency and dense mesh breakdown. Consider the dual projectors [10] $\mathbf{P}^{\Lambda} = \boldsymbol{\Lambda}\left(\boldsymbol{\Lambda}^T\boldsymbol{\Lambda}\right)^{+}\boldsymbol{\Lambda}^T$ and $\mathbf{P}^{\Sigma H} = \mathbf{I} - \mathbf{P}^{\Lambda}$.

Also, define the matrix $\mathbf{T}_s$ as the operator $T_s$ discretized with the BC basis function: $(\mathbf{T}_s)_{i,j} = \langle \hat{\boldsymbol{n}} \times \boldsymbol{g}_i, T_s \boldsymbol{g}_j \rangle$ Define also the following rescaling operator

$$\mathbf{M}_3 = \mathbf{P}^{\Sigma H} \mathbf{T}_s \mathbf{P}^{\Sigma H} + \mathbf{P}^{\Lambda}. \tag{34}$$

This rescaling operator is designed to provide a regularization of negative order $T_s$ only where needed (the hypersingular operator), while the part of the spectrum which is regular already (due to the presence of the identity in S, see considerations in the previous section) is not further regularized. Finally, the regularized low-frequency and dense grid stable IBC-EFIE we propose reads

$$\mathbf{M}_3 \mathbf{G}_{mix}^{-1} \mathbf{M}_1 \mathbf{S} \mathbf{M}_2 \mathsf{Y} = \mathbf{M}_3 \mathbf{G}_{mix}^{-1} \mathbf{M}_1 \mathsf{V} \tag{35}$$

The reader should notice that the inverse of the mix-Gram matrix $\mathbf{G}_{mix}^{-1}$ is used to link the RWG and BC basis functions.

### B. Properties of the formulation

The low frequency properties of the formulation in (35) are unchanged by the additional presence of the rescaling operator $\mathbf{M}_3$, so that the analysis in subsection III-A applies here unaltered. Regarding the dense grid behavior of the equation, the rationale behind equation (35) can be further understood by using spherical harmonics. In fact the continuous counterpart $M_3$ of the matrix $\mathbf{M}_3$ has the following spherical harmonics mappings

$$M_3(\boldsymbol{U}) = T_s(\boldsymbol{U}) = \left[ \underset{l \to \infty}{O} \left( \frac{1}{l} \right) \right] \boldsymbol{X} \tag{36}$$

$$M_3(\boldsymbol{X}) = \boldsymbol{U}. \tag{37}$$

The above mappings, combined with (31) and (32), results in a conditioning for (35) which is independent on both frequency and $h$.

## V. Implementation Related Details

At extremely low frequency, some precautions must be taken to avoid numerical cancellation in the computation of (35) right hand side. Especially when computing $\mathsf{V}$ for a plane wave $E^i = e^{ik\hat{\boldsymbol{r}} \cdot r^i}$, the solenoidal part $\mathsf{V}_{ext}$ should be computed using the extracted form $e^{ik\hat{\boldsymbol{r}} \cdot r^i} - 1$.

$$\mathsf{V}_1 = \frac{1}{ikd} \mathbf{M}_3 \mathbf{G}_{mix}^{-1} \mathbf{P}^{\Lambda H} \mathsf{V}_{ext} \tag{38}$$

$$\mathsf{V}_2 = \mathbf{M}_3 \mathbf{G}_{mix}^{-1} \mathbf{P}^{\Sigma} \mathsf{V}. \tag{39}$$

Also, to compute the system matrix in (35) and avoid numerical cancellation at extremely low frequency, the properties $\mathbf{T}_h \mathbf{P}^{\Lambda H} = \mathbf{P}^{\Lambda H} \mathbf{T}_h = 0$ and $\mathbf{P}^{\Sigma} \mathbf{T}_h \mathbf{P}^{\Sigma} = \mathbf{T}_h$ [10] must be explicitly enforced, i.e. defining

$$\mathbf{A}_1 = -\mathbf{M}_3 \mathbf{G}_{mix}^{-1} \mathbf{M}_1 z (\mathbf{K} + \frac{\mathbf{G}_{mix}}{2}) \mathbf{G}_{mix}^{-1} \mathbf{G} \mathbf{M}_2 \tag{40}$$

$$\mathbf{A}_2 = \eta i k \mathbf{M}_3 \mathbf{G}_{mix}^{-1} \mathbf{M}_1 \mathbf{T}_s \mathbf{M}_2 \tag{41}$$

$$\mathbf{A}_3 = b \eta \mathbf{M}_3 \mathbf{G}_{mix}^{-1} \mathbf{T}_h \tag{42}$$

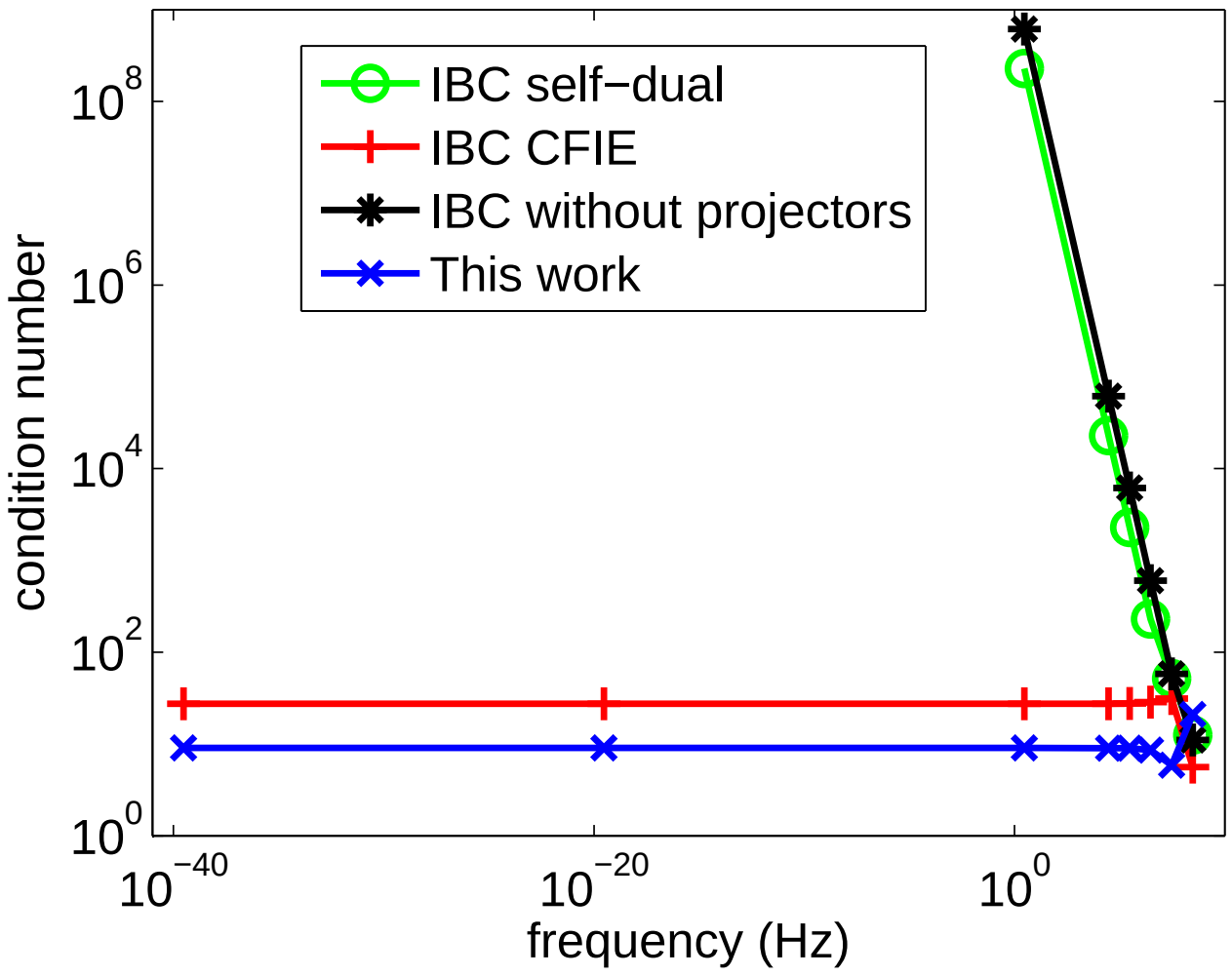

Fig. 1. Condition number as a function of the frequency on the unit sphere ($z = (0.7 + 0.6i)\eta$, $h = 0.15$m)

(35) becomes

$$(\mathbf{A}_1 + \mathbf{A}_2 + \mathbf{A}_3)\mathsf{Y} = \mathsf{V}_1 + \mathsf{V}_2. \tag{43}$$

The solenoidal part and the irrotational part of the currents are retrieved with :

$$\mathsf{j}_s = \frac{\eta i k d}{z + i\eta k a}\mathbf{P}^{\Lambda H}\mathsf{Y} \tag{44}$$

$$\mathsf{j}_{ns} = ikb\mathbf{P}^{\Sigma}\mathsf{Y} \tag{45}$$

$$\mathsf{m}_s = -z\mathbf{G}_{mix}^{-1}\mathbf{G}\mathsf{j}_{ns} \tag{46}$$

$$\mathsf{m}_{ns} = -z\mathbf{G}_{mix}^{-1}\mathbf{G}\mathsf{j}_s. \tag{47}$$

Again, at extremely low-frequency the scattered field of the solenoidal part of the currents should be computed using the extracted form of the Green's function in the integration (see above) to avoid numerical cancellation.

## VI. Numerical Results

A first set of tests for the new formulation has been conducted on the unit sphere, for which an analytic solution is available. In Fig. 1, the condition number is plotted against the frequency on a unit sphere. The impedance is kept constant and equal to $z = (0.7 + 0.6i)\eta$. The mesh parameter is set equal to $h = 0.15$m. Our formulation is compared with other two well-established formulations ([4] and [6]). The results clearly confirm that our scheme is immune from the low-frequency breakdown.

The behavior of the conditioning as a function of the mesh density is tested in Fig. 2 where the condition number is plotted against the average edge size on the unit sphere. The impedance and the frequency are set equal to $z = (0.7 + 0.6i)\eta$ and $f = 60$MHz for all simulations. The condition number remains constant when the discretization increases showing that our equation is immune from the dense mesh breakdown.

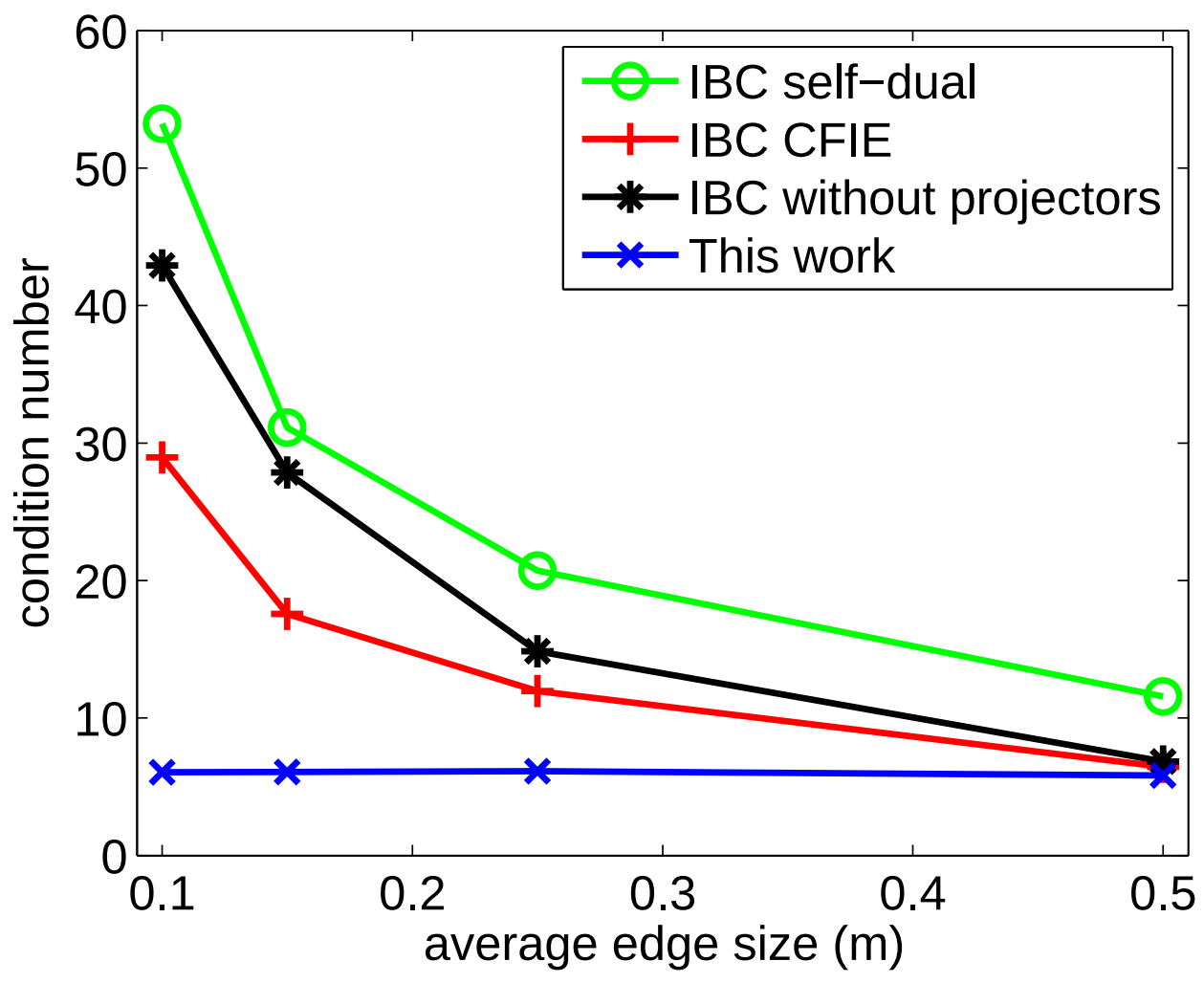

Fig. 2. Condition number as a function of the mesh size on the unit sphere mesh ($z = (0.7 + 0.6i)\eta$, $f = 60$MHz)

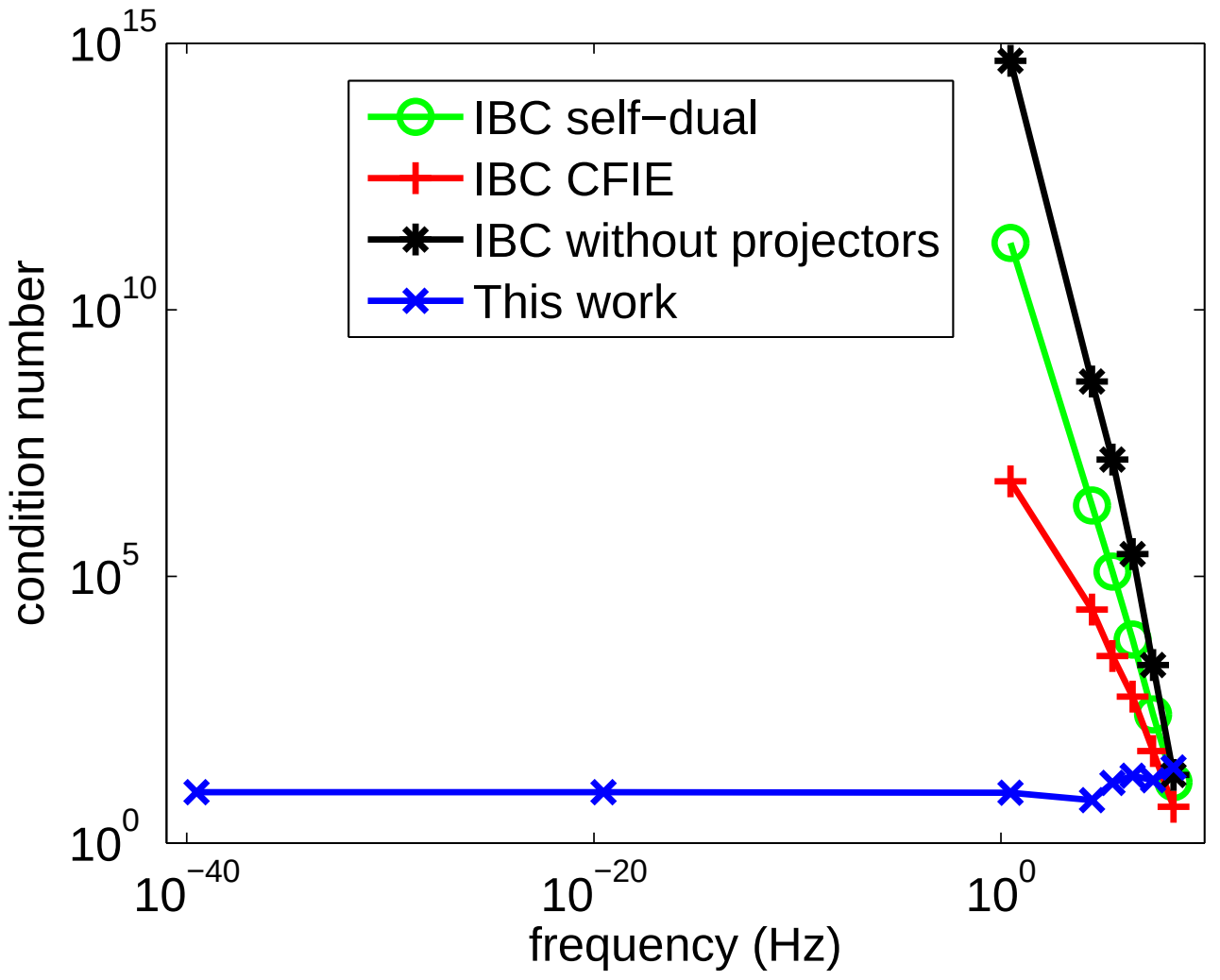

Fig. 3. Condition number as a function of the frequency on the unit sphere ($z = 1.0268 \times 10^{-7}\sqrt{\omega}(1 - i)$, $h = 0.15$m)

To check that our scheme is immune from the low-frequency breakdown also when the impedance is a function of the frequency, Fig.3 plots the condition number of the equation for an impedance equal to $z = 1.0268 \times 10^{-7}(1 - i)\sqrt{\omega}$. Again, the condition number remains constant for decreasing frequencies.

The matching of the solution obtained with our equation with the analytic solution is verified in Fig.s 4 and 5 which compare the RCS obtained with several IBC formulations and the analytic one obtained via a Mie series. The impedance is set equal $z = (0.7 + 0.6i)\eta$ and the frequency is $f = 1$MHz. A matching and convergent solution is clearly evident.

The performance of the formulation on a realistic case scenarios are assessed by simulating the RCS behavior of the F-117 Nighthawk model in Fig. VI with and without coating ($z = \eta$ and $z = 0$ respectively). The electric and magnetic current magnitudes for frontal incidence are represented in Fig. 7 and 8, respectively, while the RCS for different incidence angles are shown in Fig. 9. The effect of the coating is clearly visible as an RCS reduction, further validating the correctness of our formulation.

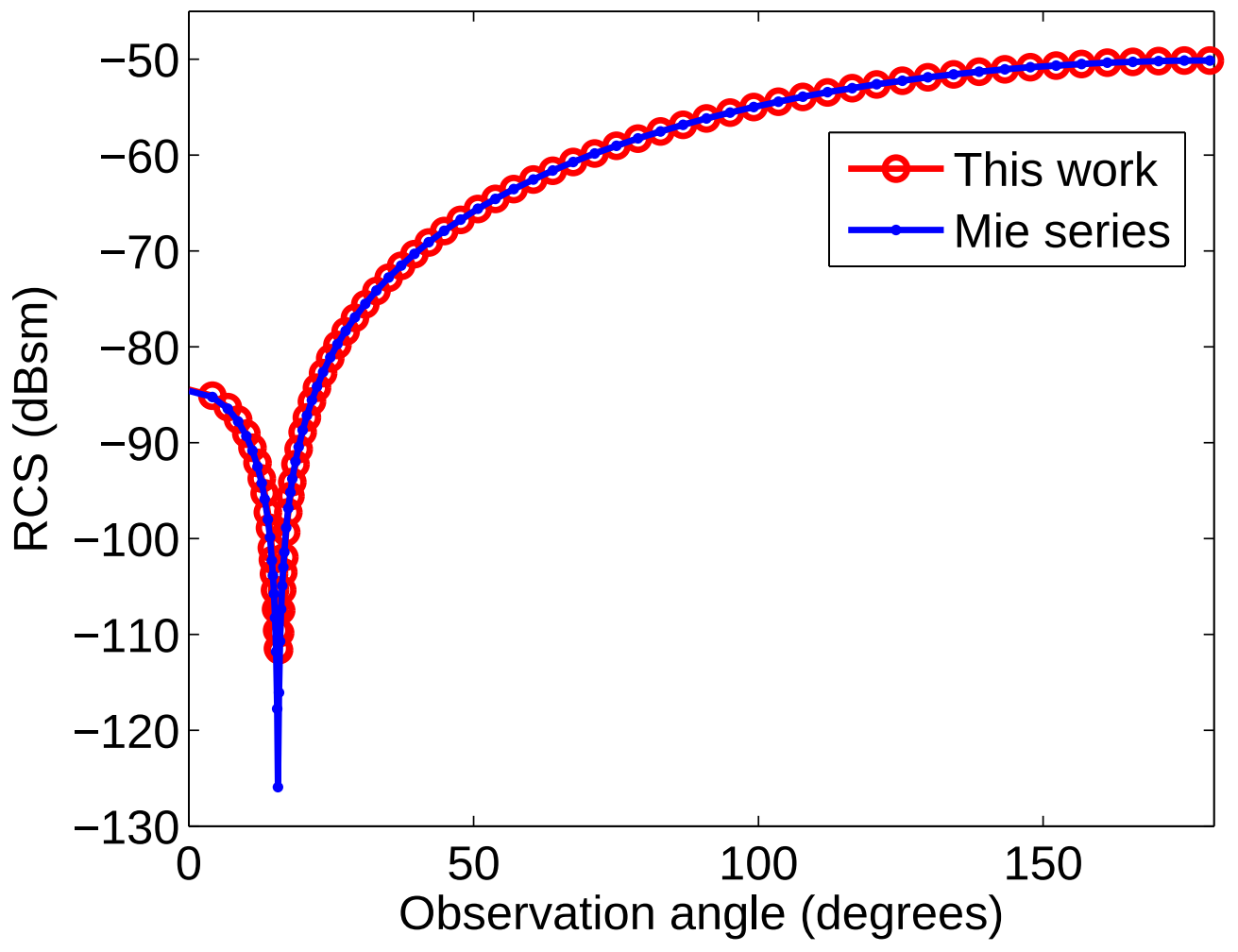

Fig. 4. RCS of a unit sphere mesh ($z = (0.8 + 0.6i)\eta$, $f = 1$MHz, $h = 0.15$m)

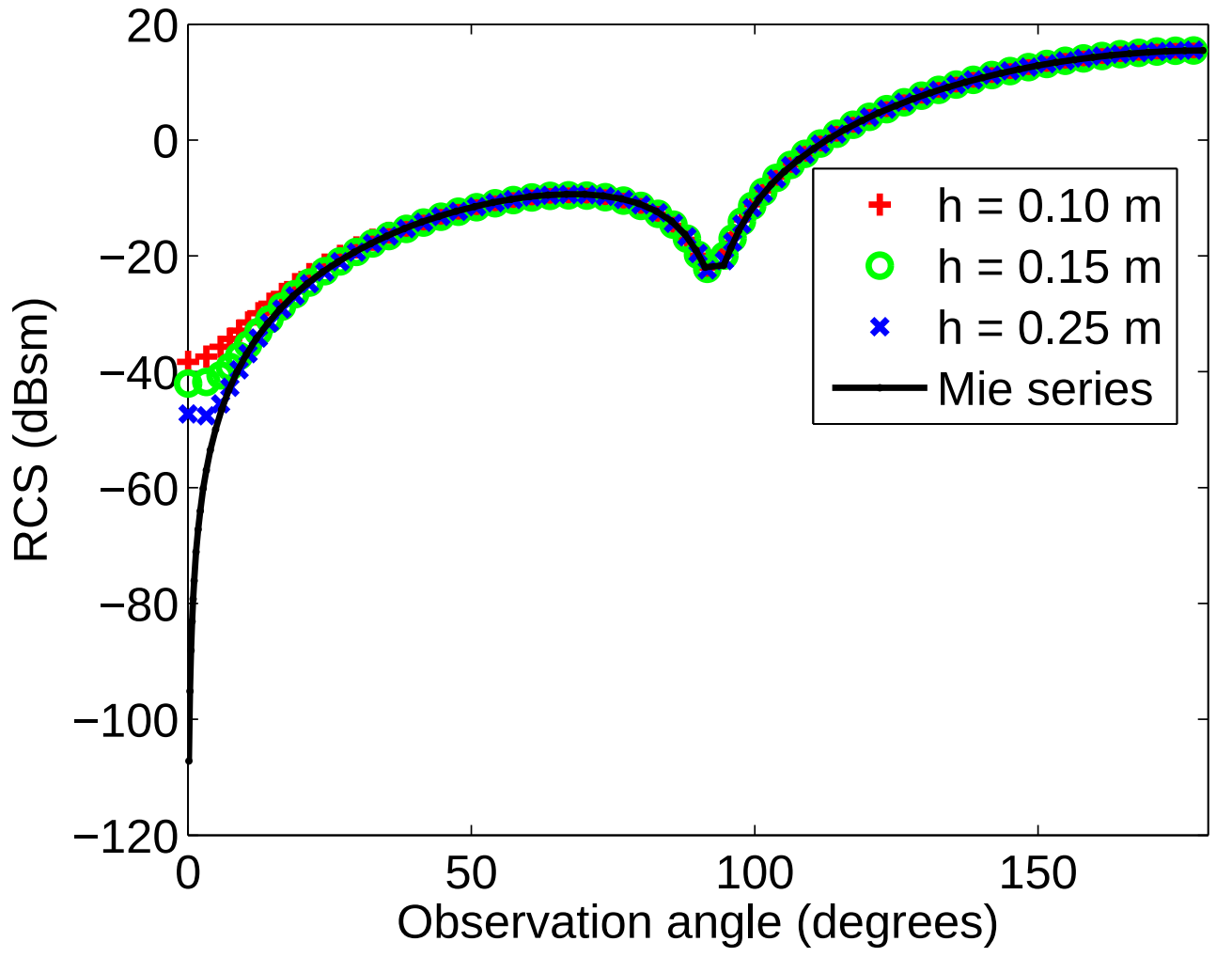

Fig. 5. RCS of a unit sphere mesh ($z = \eta$, $f = 100$ MHz)

## VII. Conclusion

In this contribution, a discretization of the IBC-EFIE was introduced, together with a left-right preconditioner that results in a linear system with condition number that remains bounded for arbitrarily low frequencies and as the mesh density is increased. The preconditioner is purely multiplicative and is based on on one hand a Helmholtz decomposition and rescaling of the current coefficient space and on the other hand on a Calderon type regularization to regularize the unbounded branch in the singular value spectrum of the single layer potential. At the same time as regularizing the condition number of the system, the method guarantees that no current cancellation occurs in the solution vector and in the right hand side of the system. Numerical results demonstrate the efficiency of the proposed methodology and this both on benchmark examples and in real life scenarios.

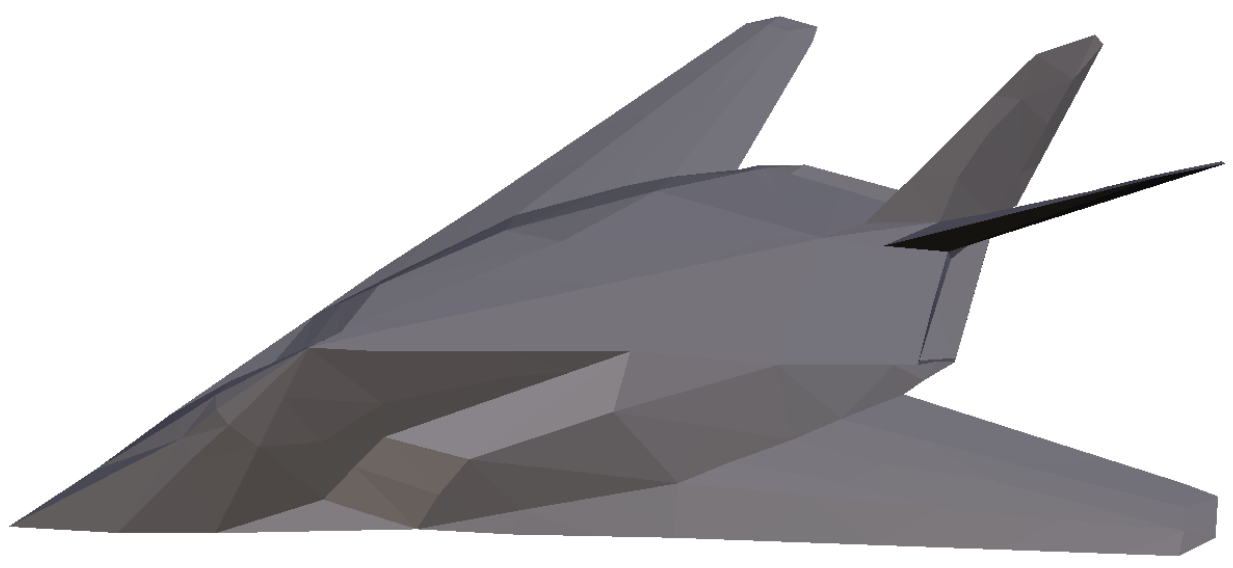

Fig. 6. F-117 Nighthawk model

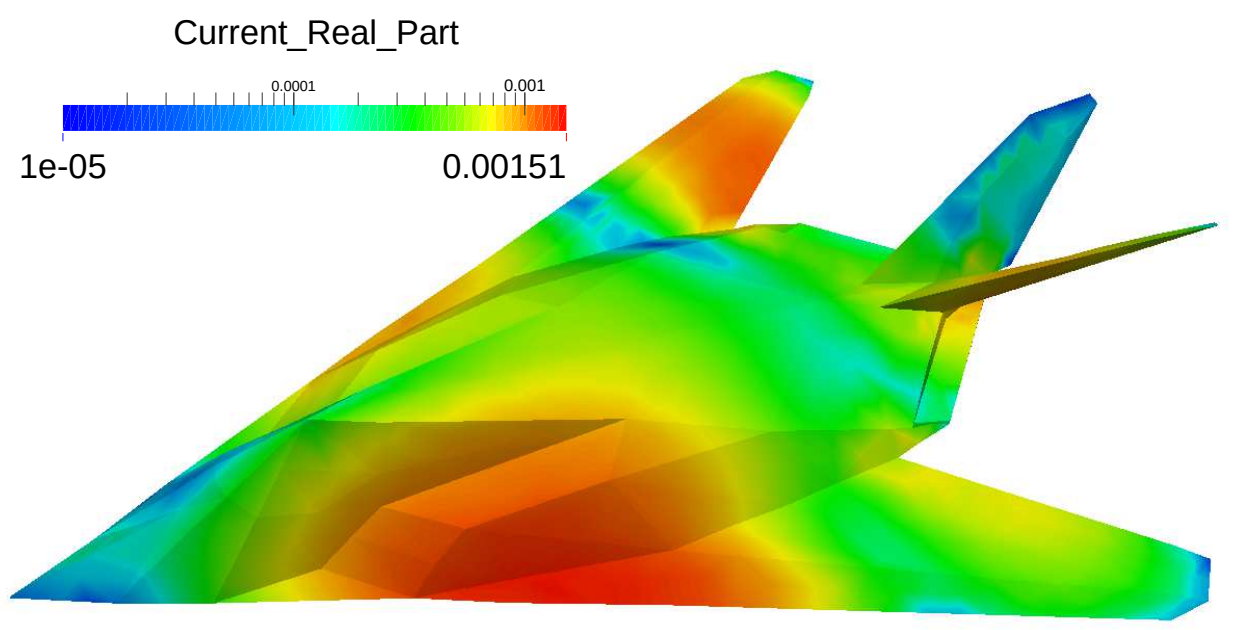

Fig. 7. Real part of the current J ($z = (0.7 + 0.6i)\eta$, $f = 10$ MHz)

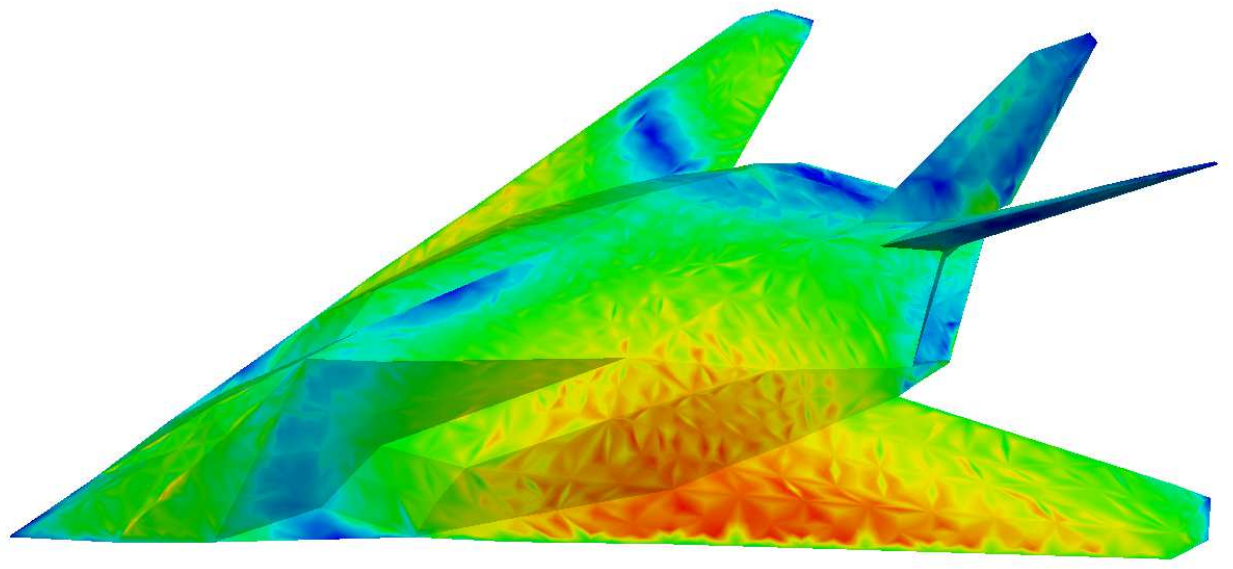

Fig. 8. Real part of the current M ($z = (0.7 + 0.6i)\eta$, $f = 10$ MHz)

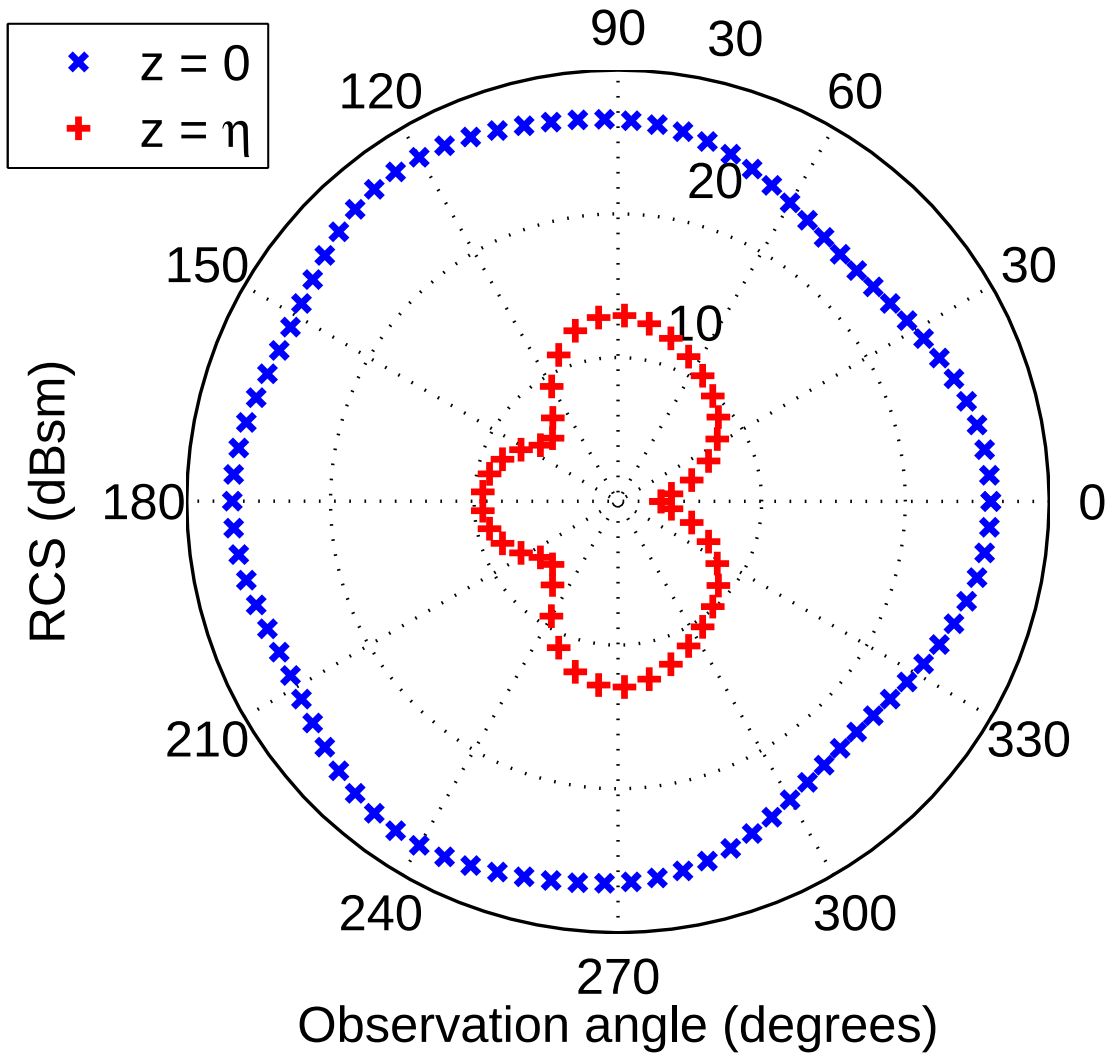

Fig. 9. RCS of a plane with ($z = 0$) and without coating ($z = \eta$) ($f = 10$ MHz)